# Application of non-uniform laxity to EDF for aperiodic tasks to improve task utilisation on multicore platforms

K Pradheep Kumar and A P Shanthi
*Department of CSE*
*Anna University, Chennai, India*
pradheepkumark@gmail.com apshanthi@annauniv.edu

***Abstract*-This paper proposes a new scheduler applying the concept of non-uniform laxity to Earliest Deadline First (EDF) approach for aperiodic tasks. This scheduler improves task utilisation (Execution time / deadline) and also increases the number of tasks that are being scheduled. Laxity is a measure of the spare time permitted for the task before it misses its deadline, and is computed using the expression (deadline - (current time + execution time)). Weight decides the priority of the task and is defined by the expression ((quantum slice time / allocated time)*total core time for the task). Quantum slice time is the time actually used, allocated time is the time allocated by the scheduler, and total core time is the time actually reserved by the core for execution of one quantum of the task. Non-uniform laxity enables scheduling of tasks that have higher priority before the normal execution of other tasks and is computed by multiplying the weight of the task with its laxity. The algorithm presented in the paper has been simulated on Cheddar, a real time scheduling tool and also on SESC, an architectural simulator for multicore platforms. The algorithm has been tested varying random task sets upto 5000 and number of cores upto 100. The algorithm improves task utilisation by 35% and increases the number of tasks scheduled by 36%, compared to conventional EDF.**

## I INTRODUCTION

In multicore platforms, when normal scheduling algorithms like EDF (Earliest Deadline First) are used, tasks are scheduled based on the actual allocation of resources. EDF schedules tasks in the order of increasing deadlines. In such approaches, task utilisation is poor, as there is no room for scheduling newly arriving tasks. This results in some tasks missing their deadline due to exhaustion of resources. Hence new techniques have to be incorporated to schedule newly arriving tasks as explained by Burchard et al in [1].

To overcome the above mentioned drawbacks of task utilisation, priority is used as one of the criterion for scheduling, as explained by Moir and Ramamurthy in [2]. Each task is assigned a weight that is based on the priority of the task.

The scheduling is carried out, according to this weight. Fairness is a criterion by which scheduling is decided. Fairness is granted to the task in proportion to the weight, and resources are allocated proportionate to fairness. This strategy is termed as proportionate fairness or pfairness. Here, tasks are scheduled by utilizing only a small portion of the allocated resource for each task. The remaining resource for each task is cumulated and later utilized for scheduling newly arriving tasks.

This strategy restricts the migration overhead around quantum bounds ranging from - $(\gamma + c\alpha/2)$ to $+(\gamma + c\alpha/2)$, where $\gamma$ is the queue overhead, c is the average number of tasks in the queue and $\alpha$ is the start time for execution of first task in the execution queue as explained in [3] by Calandrino et al.

On applying the concept of proportionate fairness, only a small quantum of resources is utilized. The remaining of the allocated resources can be cumulated as laxity for use for newly arriving tasks. Laxity is a measure to identify the slack time and the urgency for the execution of tasks.

Laxity as applied to scheduling is of three categories namely zero laxity, least laxity first and uniform laxity.

In the zero laxity approach, a task is dispatched for execution when laxity becomes zero. In this approach, the scheduler does not act until laxity becomes zero. The overhead imposed on the scheduler increases rapidly, as a number of simultaneous decisions need to be taken if multiple tasks arrive with zero laxity. For a two core system under zero laxity condition, the task utilisation is modified by a factor of $1.5 + |u_{max} - 0.5|$ as discussed by Wen et al in [4]. Tasks can be scheduled on a two core system only if the task utilisation is less than or equal to (z+1) / 2 as explained by Piao et al in [5]. By applying a pure zero laxity condition to EDF scheduling strategy, task sets are not schedulable when U(T) > z * (1- (1/e)), where z is the number of cores and e is the Euler's number as explained by Yi-Hsiung Chao et al in [6]. Hence U(T) is always <( z * (1- (1/e))) for tasks to be always schedulable by the laxity condition.

In the least laxity approach, tasks are sorted in order of ascending laxity and tasks with least laxity are executed first. The main drawback in this approach is that the priority gets revised when tasks newly arrive, and the earlier laxity computation gets modified with respect to the revised priority as explained by Baker in [7].

While using uniform laxity, laxity is computed for tasks initially assigned on that core and for those that arrive later, ensuring equal priority for all tasks. When priorities vary, this strategy tends to break down as explained by Dertouzos and Mok in [8]. However, uniform laxity does not yield an improvement in task schedulability as the priority is treated as unity and there is possibility of over or under utilisation of resources allocated without recalculation of weights.

In this paper, a new approach namely non-uniform laxity is applied to EDF, wherein laxity is recomputed taking revised priority into account for newly arriving tasks. This facilitates scheduling tasks in proportion to the resources actually needed for execution.

The algorithm also ensures the scheduling constraint U(T) < Z* (1- (1/e)), where z is the number of cores, and e is Euler's number. Laxity is monitored with respect to the task utilisation based on the weight, which is a measure of priority.

The algorithm has been simulated on Cheddar, a real time scheduling tool in [9] and [10] to check the scheduling constraints and also on SESC, an architectural simulator in [11] and [12] to check the schedulability on multicore platforms. This algorithm has been tested varying random task sets upto 5000 and number of cores upto 100. The algorithm improves task utilisation by 35% and increases the number of tasks scheduled by 36%, compared to conventional EDF.

The paper is organized as follows: Section II discusses the algorithm and an example. Section III discusses the Cheddar and SESC simulation results of the algorithm in multicore platforms. Section IV concludes the work and presents the scope for future work.

## II ALGORITHM

The main components of the algorithm are explained below:

### A. *Laxity computation:*

The laxity of the task is a measure for slack time. The laxity is computed by the equation laxity = (deadline of task - (current time of task +execution time for task)).

The laxity gives an intuitive direction for task execution. The following are the three cases for laxity computation.

(i) Positive laxity: When laxity of a task is positive, the execution can be delayed, and the task can be placed in a holding queue.
(ii) Zero laxity: When laxity of a task is zero, the task needs to be dispatched immediately to the execution queue, failure to do so might lead the task to miss its deadline.
(iii) Negative laxity: When laxity of a task is negative, the task has missed its deadline and the same is discarded from the holding queue. Normally this condition is eliminated by the algorithm.

### B. *Weight computation:*

Each task is assigned a weight in proportion to its respective priority. The weight of the task is computed based on the quantum slice time used for executing the task. The weight of the task is computed by the equation
Weight of task = ((Quantum slice time used / Actual time allocated by the scheduler)*Total time allocated by each for each task).

The quantum slice time is the time actually used for execution of the task.

The actual time allocated is the time allocated initially by the scheduler for all tasks. This indicates the complete resources held by the scheduler.

The total core time for the task is the time allocated by the core for continuous execution of a single quantum of the task.

### C. *Non-uniform laxity:*

The non-uniform laxity of each task is computed by the equation
Non-uniform laxity = (laxity of the task * weight of the task)

This modified laxity facilitates scheduling of tasks that arise due to preemption or tasks which arrive later.

### D. *Task utilisation:*

The task utilisation is computed by the equation
Original task utilisation = execution time of task / deadline of the task, for aperiodic tasks.
Modification factor for task utilisation =1.5+ (|maximum task utilisation for task set – 0.5 |)
Modified task utilisation is computed by the equation
Modified task utilisation = (Modification factor * original task utilisation).

The main steps of the algorithm are outlined below:

- Start scheduling tasks using EDF.
- Define two queues namely holding (H) and execution queue (X) for performing operations on tasks.
- Define the factor L=(z*(1-(1/e)));
- Compute original task utilisation task utilisation=(execution time/ deadline);
- Compute modification factor =(1.5+(|maximum task utilisation-0.5|));
- Modify task utilisation using equation modified task utilisation = (modification factor *original task utilisation);
  // Check for 2 cores
  if (modified task utilisation <((z+1)/2)), append task to execution queue.
- Execute tasks on the respective cores.
  // For more than 2 cores
  if (modified task utilisation <2), then append task to holding queue H.
  If (modified task utilisation >=2), then append task to execution queue X.
- Execute the tasks on the respective cores.
  //Compute the weight of the tasks.
- Weight of task = ((quantum slice time/total allocated time by scheduler)*total core time for execution of one quantum of the particular task)
  //Compute laxity and non-uniform laxity of the tasks.
- Laxity of the task = (deadline-(current time + execution time))
- Non-uniform laxity of the task = (laxity of the task * weight of the task).
  //Check non-uniform laxity conditions to remove task from holding queue and append to execution queue.
- If (non-uniform laxity >0), then append task to holding queue.
- If ((non-uniform laxity =0) & (modified task utilisation <2)), then urgently remove task from holding queue and append to execution queue.
- Execute task on same core.
- If ((non-uniform laxity =0) & (modified task utilisation >=2)), then remove task from holding queue and append to execution queue.

- If (modified task utilisation <2), then execute task on next core.
- If ((non-uniform laxity<0) & (modified task utilisation > L)), then the task has missed its deadline.
- Discard the task from the holding queue H.
- End the procedure.

Actual algorithm is furnished below:

//Algorithm for task scheduling incorporating non-uniform laxity //

Global variables
exec : array of execution times $task_i$[i=1…n];
dline : array of deadlines of $task_i$[i=1…n];

Arrays:

V: array of rationalized utilisation $task_i$ [i=1…n];
U: array of task utilisations of $task_i$ [i=1…n] exec/dline;
U2: array of task utilisations of $task_i$ [i=1…n] exec/dline;
$U_{max}$: maximum task utilisation computed from task set under discussion;
U1: Constant computed for task utilisation of $task_i$ [i=1…n] based on $1.5+|u_{max}-0.5|$;
e: Euler's number (type real);
L: constant (type real);
start : array of tasks [i=1…n] scheduled on cores[z=1…m];
cur : array of current times of $task_i$[i=1…n];
quant : Quantum slice of time used for $tasks_i$[i=1…n];
total : total allocated time of $tasks_i$[i=1…n];
ctot : Core total time for $tasks_i$[i=1…n];
lax : array of laxities of $task_i$[i=1…n];
nonunilax: array of non-uniform laxities of $task_i$[i=1…n];
W: weight of $tasks_i$[i=1…n];
X: Execution queue;
H: Holding queue ;
> identifier for current core
$core_z$ : z varies from 1…m initially 1;
> identifier for current task
$task_i$ : i varies from 1…n initially 1;
local variables
m : number of cores (type int);
n: number of tasks (type int);

// Normal scheduling based on EDF
1. Start scheduling based on EDF;
// Computation of task utilisation
2. H=0;
3. L=(z*(1-(1/e)));
4. for $core_z$ (z=1 to m) do
5. for $task_i$ (i=1 to n) do
// compute modified task utilisation
6 U[$task_i$]=exec/dline;
7. U1=(1.5+(|$U_{max}$-0.5|));
8. U2[$task_i$]= (U1*U[$task_i$]);
// check condition for two cores
9 if ((z=2) & (U2[$task_i$]<((z+1)/2))) then //put this task in execution queue
10. X=X +exec[$task_i$];
// Execute task on the respective core
11. start[$core_z$]=start[$core_z$]+exec[$task_i$];
fi
// check condition for more than two cores
12. if(z>2) then
13. if(U2[$task_i$]<2) then
//put the task in holding queue
14. H=H +exec[$task_i$];
else
15 if(U2[$task_i$] =2 | U2[$task_i$] >2) then
//put the task in execution queue
16. X=X +exec[$task_i$];
// execute task on respective core
17. start[$core_z$]=start[$core_z$]+exec[$task_i$];
18. Execute the next task;
fi
fi
fi
od
od
// Compute the weight
19. for $task_i$ (i=1 to n) do
20. W[$task_i$]=(quant[$task_i$]/total[$task_i$])*ctot[$task_i$];
od
// compute Laxity and non-uniform laxity
21. for $task_i$(i=1 to n) do
22. lax[$task_i$]=(dline[$task_i$]–(cur[$task_i$] + exec[$task_i$]));
23. nonunilax[$task_i$]=(lax[$task_i$] * W[$task_i$]);
od
// check non-uniform laxity conditions
24. X=0;
25. for $core_z$(z=1 to m) do
26. for $task_i$(i=1 to n) do
// operation for positive non-uniform laxity
27. if(nonunilax[$task_i$]>0) then
28. H=H+exec[$task_i$];
fi
// operation for zero non-uniform laxity
29. if ((nonunilax[$task_i$])=0 & (U2[$task_i$]>=2)) then
30. H=H -exec[$task_i$];
31. X=X +exec[$task_i$];
// Execute task on the same core
32. start[$core_z$]=start[$core_z$]+exec[$task_i$];
fi
// operation for modified task utilisation <2
33. if (U2[$task_i$]< 2) then
34. start[$core_z$]=start[$core_z$]+exec[$task_i$];
fi

// operation for negative non-uniform laxity
35. if ((nonunilax[$task_i$])<0& (U2[$task_i$] >L) then
36. H=H -exec[$task_i$];
37. Display message "Task missed dead line"
fi
od
od
38. End the procedure.

As a proof of concept, the working of this algorithm for an example task set is presented below:

*A Example*

A task set comprising of six tasks is assumed to be scheduled on four cores. The tasks are intra-sporadic. (Tasks can be scheduled in a random manner with respect to a particular core.). Table I indicates the arrival time, execution time, current time, deadline, utilized execution time and weight of each task.

**TABLE I** CURRENT TIME, DEADLINE , QUANTUM SLICE TIME ALLOCATED EXECUTION TIME CORE TOTAL TIME FOR TASKS

| Tasks | Current time (ct) | Dead line (d) | Quantum slice time (Quant) | Allocated Execution time (exec) | Core total time (ctot) alotted |
|---|---|---|---|---|---|
| T1 | 0 | 125 | 10 | 80 | 7 |
| T2 | 0 | 140 | 15 | 100 | 6 |
| T3 | 75 | 200 | 20 | 120 | 5 |
| T4 | 100 | 260 | 30 | 140 | 4 |
| T6 | 125 | 300 | 25 | 160 | 5 |
| T5 | 250 | 500 | 28 | 210 | 6 |

The weights for the tasks are calculated as a proportion based on the actual used time for execution. The weight for each task is computed, based on the relation

weight=((quantum slice time / Allocated time for each task by the scheduler)*Total time on the core allocated for each task).

For example, weight of task 2 is ((15/100)*6) (vide table I, row 2) which is 0.9 (vide table II, row 2). Based on this, the non-uniform laxity of the task is computed. Laxity = deadline - (current time + execution time).

Laxity for the task T2 is (140 - (100 + 0)) (vide table I, row 2 ) = 40 units (vide table II, row 2). Task T2 uses (40*0.9) which is 36 units (vide table II, row 2) which is the non-uniform laxity. This is the time actually utilized by the task for execution out of the totally allocated time of 100 units (vide table I, row 2).

TABLE II
WEIGHT, LAXITY, NON-UNIFORM LAXITY AND TASK UTILISATION FOR EACH TASK

| Tasks | Weight Wt = (Quant / exec) * ctot | laxity lax = d - (c+e) | Non-uniform laxity nlax = (lax * wt) | Task utilisation(exec /d) | Modified task utilisation (nlax/d) |
|---|---|---|---|---|---|
| T1 | 0.88 | 45 | 39.38 | 0.64 | 0.32 |
| T2 | 0.90 | 40 | 36.00 | 0.71 | 0.26 |
| T3 | 0.83 | 5 | 4.17 | 0.60 | 0.02 |
| T4 | 0.86 | 20 | 17.14 | 0.54 | 0.07 |
| T6 | 0.78 | 15 | 11.72 | 0.53 | 0.04 |
| T5 | 0.80 | 40 | 32.00 | 0.42 | 0.06 |

The algorithm is simulated using Cheddar tool. Cheddar facilitates monitoring task utilisation and corresponding laxity. Cheddar also identifies tasks that have to be dispatched urgently to execution queue.

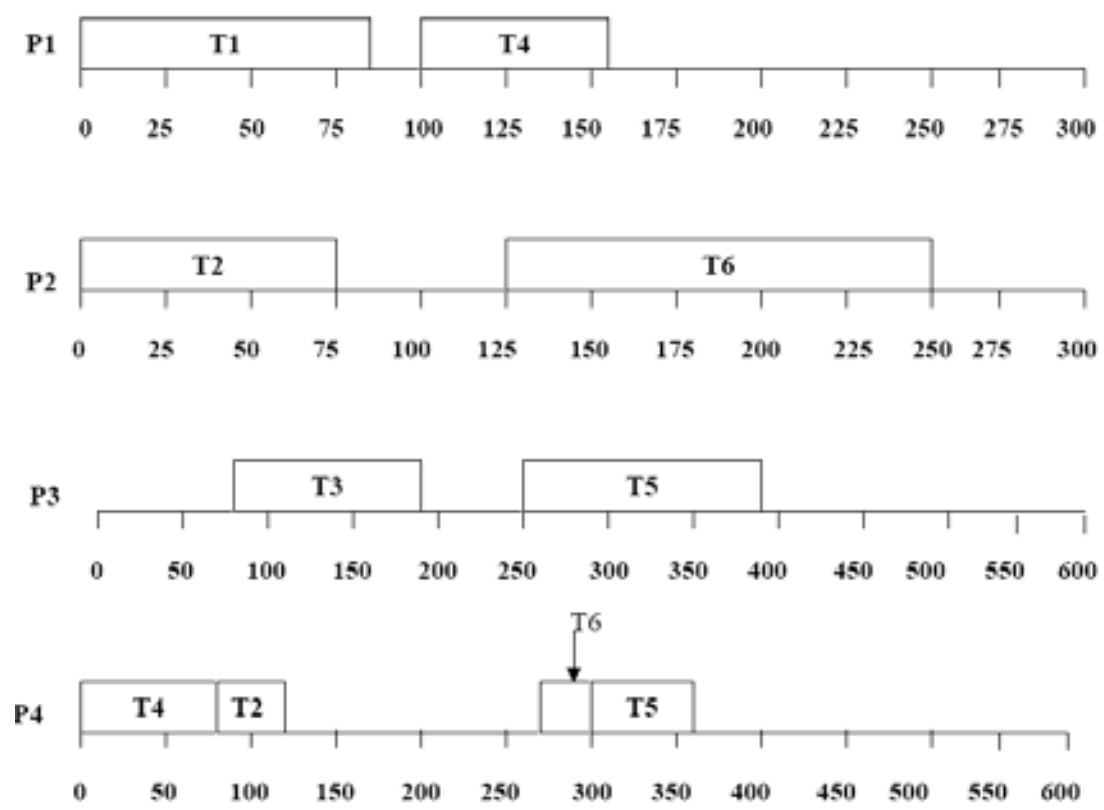

Fig. 1. Conventional EDF Schedule without slack time measures.

From Fig 1, the task utilisation of individual cores can be computed. For example, task utilisation for T1 is 0.64 (80/125 vide table I) and task utilisation for T4 is 0.23 (dividing 59.8 by 260 vide fig 1 and table I). Task utilisation on core 1 is 0.87 (obtained by adding 0.64 and 0.23). Similarly task utilisation for core 2, core 3 and core 4 are 0.94, 0.87 and 0.66 respectively. The task utilisation for conventional EDF for aperiodic tasks is computed based on execution time and deadline. The execution cost / deadline is the task utilisation. For example, for task 1, utilisation is 80/125 (vide table I) i.e, 0.64 (vide table II). Similarly the other task utilisations can be computed and then the task utilisations on each core are summed up. This total task utilisation on each core does not exceed 1, as is obvious from Fig 1. The average task utilisation on core 1 is 0.44 (dividing 0.87 which is total task utilisation on core 1, by 2 which is the number of tasks scheduled on core 1). Similarly average task utilisations for core 2, core 3 and core 4 are 0.47, 0.43 and 0.16 respectively. The average task utilisations of the task without including the laxity factor comprising of task utilisations 0.44, 0.47,0.43 and 0.16 is 0.38. This value is the task utilisation of the entire task set for conventional EDF. This approach completely exhausts all the resources allocated by the scheduler initially, and newly arriving tasks can not be accommodated resulting in infeasible schedule.

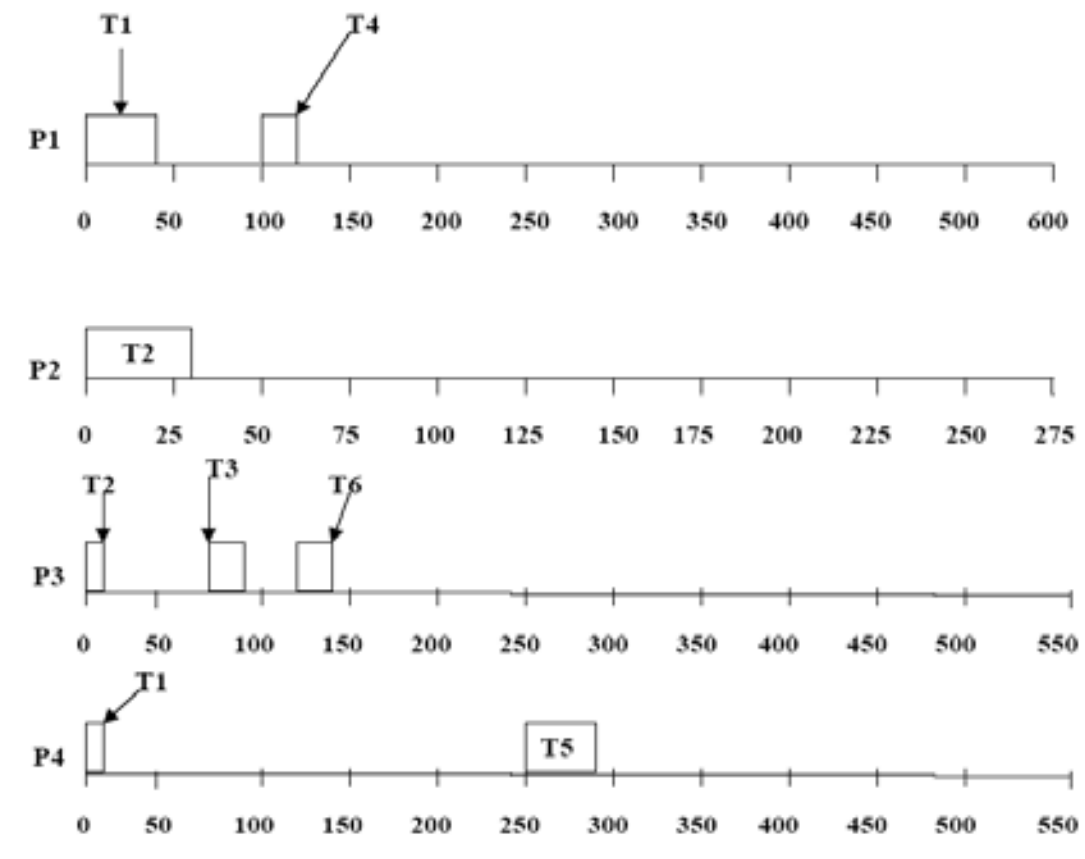

Fig. 2. Modified EDF Schedule incorporating non-uniform laxity.

Fig 2. shows the schedule after incorporating non-uniform laxity approach. Task utilisations are modified by the factor 1.5 + (| $u_{max}$ – 0.5 |). The maximum of the task set is then computed. From table II $u_{max}$ = 0.32 (vide table II, row 1). $u_{max}$ is computed as the maximum task utilisation from the modified task utilisations (vide table II). The modification factor for task utilisation is then computed as 1.5+(|0.32-0.5|) i.e., 1.5+0.18=1.68. Tasks are monitored for

the condition U(task$_i$) < (z * (1- (1/e))) so that no task misses the deadline. The value of (z*(1-(1/e))) is 2.528, where Euler's number e is 2.718 and the number of cores z is 4. The modified task utilisation is (1.68*0.71) =1.19 which is < 2.528. Task utilisation for a two core system is always less than (z+1) / 2. The task utilisation for conventional EDF for first two cores computed from fig 1, is 0.87+0.94 = 1.81 which is < 2. After incorporating non-uniform laxity, task utilisation for T1 and T4 are 0.64 and 0.54 respectively (derived from fig II). Hence the task utilisation on core 1 is 1.68*(0.64+0.54)=1.98, where 1.68 is the modification factor. Similarly for core 2, core 3 and core 4, the task utilisations are 1.19, 1.71 and 1.78 respectively. The average task utilisation on core 1 is 1.96/2=0.98 as number of tasks is 2 (vide fig 2). Similarly for core 2, core 3 and core 4 the average task utilisations are 1.19, 0.57 and 0.89 respectively. The average task utilisation for the entire task set having task utilisations 0.98,1.19,0.57 and 0.89 is 0.91. This value is the task utilisation of the task set after incorporating non-uniform laxity.

The algorithm has been simulated using Cheddar tool and the SESC simulator. The results are provided in the next section.

## III Simulation Results

### A Task Utilisation

The improvement in task utilisation has been analysed using Cheddar, a real time scheduling tool .

TABLE III
IMPROVEMENT IN TASK UTILISATION FOR NON-UNIFORM LAXITY APPROACH - SIMULATED ON CHEDDAR

| | Task utilisation = (execution time / deadline) | | |
|---|---|---|---|
| **Number of cores** | **EDF** | **Non- uniform laxity applied to EDF** | **Improvement %** |
| 4 | 0.93 | 0.99 | 6 |
| 8 | 0.91 | 0.97 | 7 |
| 12 | 0.87 | 0.95 | 9 |
| 16 | 0.83 | 0.93 | 12 |
| 20 | 0.81 | 0.92 | 14 |
| 24 | 0.78 | 0.9 | 15 |
| 28 | 0.76 | 0.89 | 17 |
| 32 | 0.73 | 0.86 | 18 |
| 36 | 0.71 | 0.85 | 20 |
| 40 | 0.68 | 0.82 | 21 |
| 44 | 0.65 | 0.79 | 22 |
| 48 | 0.59 | 0.73 | 24 |
| 52 | 0.55 | 0.71 | 29 |
| 56 | 0.52 | 0.69 | 33 |
| 60 | 0.49 | 0.67 | 37 |
| 65 | 0.47 | 0.66 | 40 |
| 70 | 0.46 | 0.65 | 41 |
| 75 | 0.43 | 0.62 | 44 |
| 80 | 0.37 | 0.59 | 59 |
| 85 | 0.35 | 0.56 | 60 |
| 90 | 0.33 | 0.53 | 61 |
| 95 | 0.29 | 0.47 | 62 |
| 100 | 0.27 | 0.45 | 67 |
| **Average** | **0.60** | **0.75** | **31** |

Table III indicates the percentage improvement in task utilisation for the algorithm proposed in this paper, compared to the conventional EDF. The average task utilisation is 0.75 for the proposed algorithm compared to 0.60 in conventional EDF. It can be seen that task utilisation improves by 31% with increase in number of cores.

5000 random task sets are simulated on 100 cores, so that an average of 50 task sets are scheduled on each core. It is seen from table III that per task utilisation on applying non-uniform laxity approach to EDF is 0.015 i.e., 0.75/50 where 0.75 is the average task utilisation and 50 is the average number of task sets per core. (Vide table III), whereas per task utilisation for conventional EDF is only 0.012 i.e., 0.60/50(Vide table III). Hence there is a significant improvement in per task utilisation.

Fig 3 compares the task utilisation for both algorithms with increase in number of cores.

It is observed that the task utilisation for EDF coupled with non-uniform laxity for 100 cores is 0.45, compared to 0.27 in conventional EDF thus improving task utilisation by 67% (vide Table III and fig 4). As such, application of non-uniform laxity significantly improves task utilisation in multicore platforms.

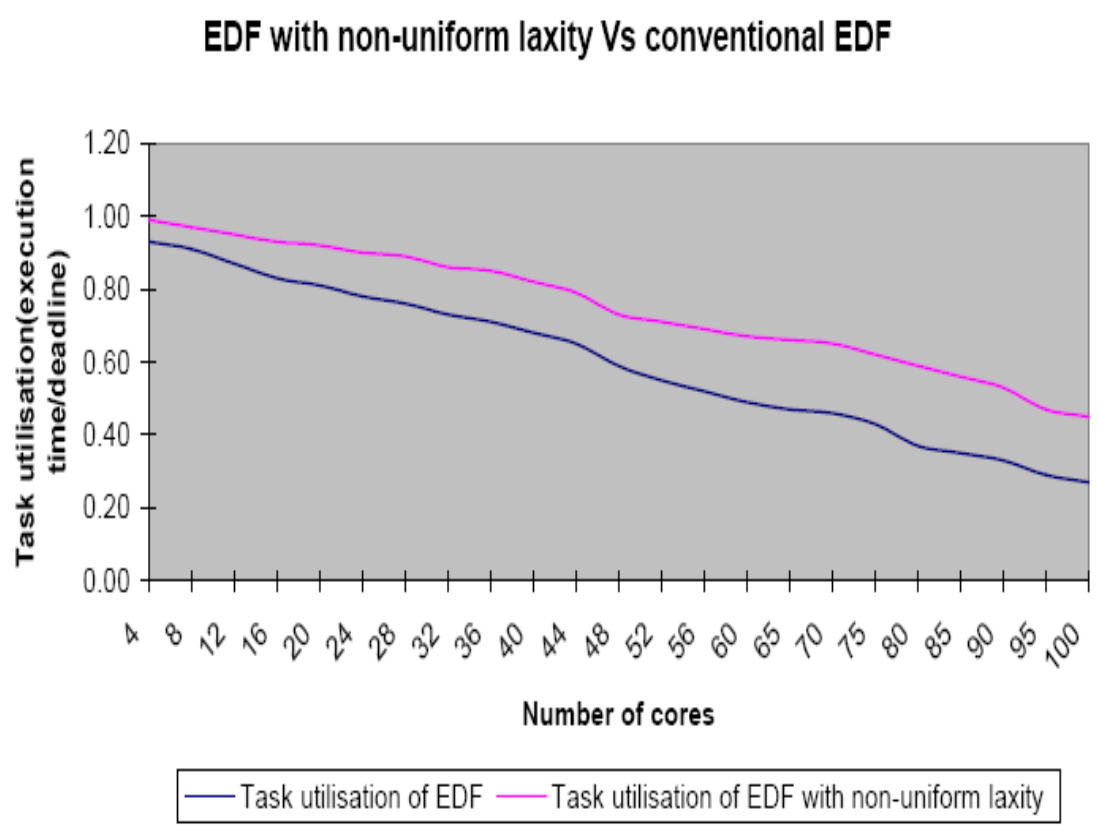

Fig. 3. Plot comparing task utilisation for both algorithms with increase in number of cores

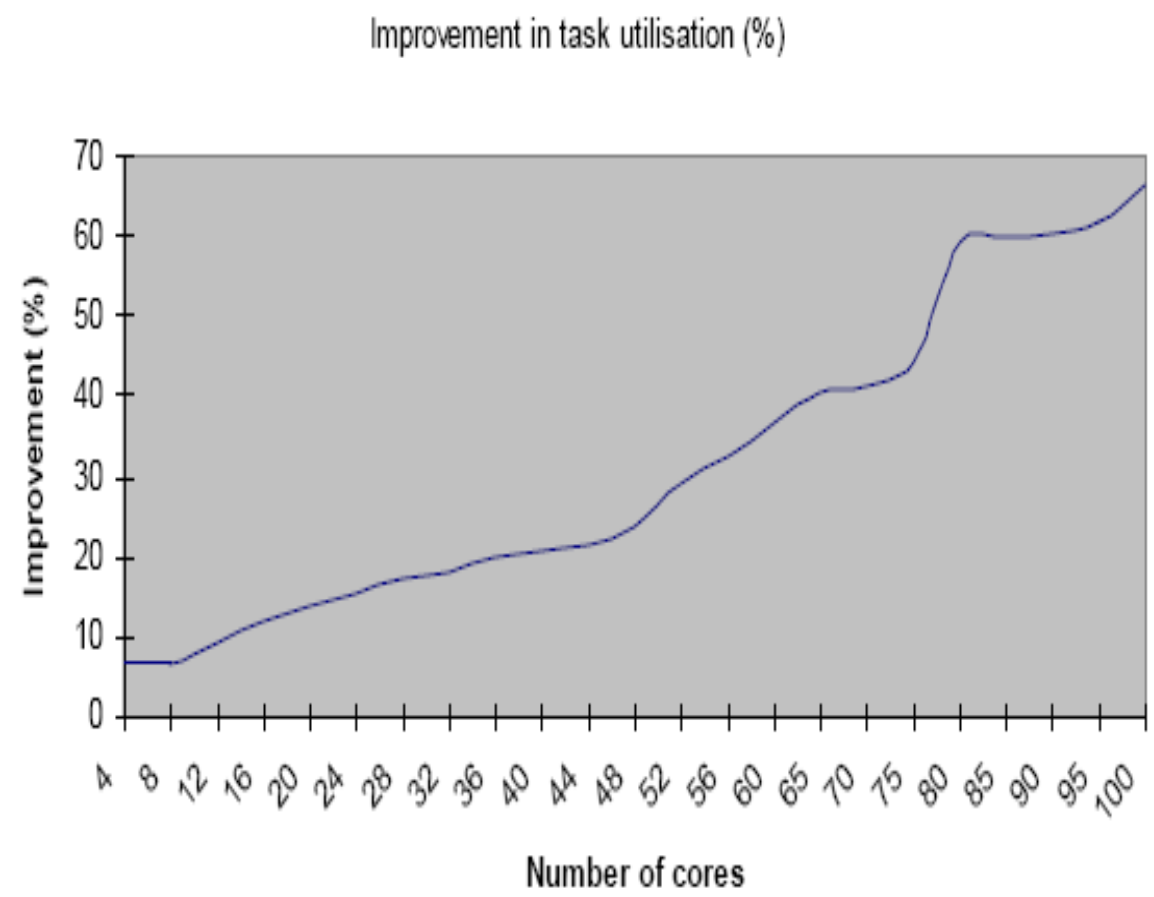

Fig. 4. Plot showing improvement in task utilisation with increase in number of cores using Cheddar

Table IV shows the task utilisation values when simulation was done using SESC, which are almost similar to the results obtained from Cheddar.

TABLE IV
IMPROVEMENT IN TASK UTILISATION FOR NON-UNIFORM LAXITY APPROACH - SIMULATED ON SESC

| Number of cores | Task utilisation = (execution time / deadline) | | Improvement % |
|---|---|---|---|
| | EDF | Non-uniform laxity applied to EDF | |
| 4 | 0.91 | 0.97 | 7 |
| 8 | 0.87 | 0.94 | 8 |
| 12 | 0.83 | 0.92 | 11 |
| 16 | 0.81 | 0.91 | 12 |
| 20 | 0.79 | 0.89 | 13 |
| 24 | 0.75 | 0.87 | 16 |
| 28 | 0.71 | 0.84 | 18 |
| 32 | 0.69 | 0.82 | 19 |
| 36 | 0.65 | 0.79 | 22 |
| 40 | 0.62 | 0.76 | 23 |
| 44 | 0.59 | 0.74 | 25 |
| 48 | 0.57 | 0.72 | 26 |
| 52 | 0.54 | 0.70 | 30 |
| 56 | 0.51 | 0.68 | 33 |
| 60 | 0.47 | 0.65 | 38 |
| 65 | 0.43 | 0.63 | 47 |
| 70 | 0.41 | 0.61 | 49 |
| 75 | 0.38 | 0.58 | 53 |
| 80 | 0.36 | 0.56 | 56 |
| 85 | 0.32 | 0.54 | 69 |
| 90 | 0.29 | 0.51 | 76 |
| 95 | 0.27 | 0.49 | 81 |
| 100 | 0.25 | 0.46 | 84 |
| **Average** | **0.57** | **0.72** | **35** |

## *B Task Schedulability*

Task schedulability is the number of tasks that are being scheduled. Task schedulability has been analysed using both Cheddar and SESC tools and results are tabulated in table V and VI respectively, which show that task schedulability increases by 36 %.

TABLE V
IMPROVEMENT IN NUMBER OF TASKS SCHEDULED FOR NON-UNIFORM LAXITY APPROACH COMPARED TO CONVENTIONAL EDF -- SIMULATED ON CHEDDAR

| Number of tasks | EDF | | Non-uniform laxity applied to EDF | | Improvement (%) |
|---|---|---|---|---|---|
| | Tasks scheduled | Tasks missed | Tasks scheduled | Tasks missed | |
| 8 | 5 | 3 | 7 | 1 | 40 |
| 15 | 8 | 7 | 11 | 4 | 38 |
| 20 | 13 | 7 | 17 | 3 | 31 |
| 30 | 18 | 12 | 25 | 5 | 39 |
| 45 | 32 | 13 | 42 | 3 | 31 |
| 60 | 40 | 20 | 55 | 5 | 38 |
| 75 | 52 | 23 | 70 | 5 | 35 |
| 80 | 55 | 25 | 75 | 5 | 36 |
| 90 | 65 | 25 | 85 | 5 | 31 |
| 100 | 71 | 29 | 95 | 5 | 34 |
| 200 | 128 | 72 | 175 | 25 | 37 |
| 500 | 378 | 122 | 465 | 35 | 23 |
| 700 | 464 | 236 | 646 | 54 | 39 |
| 900 | 682 | 218 | 845 | 55 | 24 |
| 1000 | 780 | 220 | 978 | 22 | 25 |
| 2000 | 1567 | 433 | 1934 | 66 | 23 |
| 5000 | 3468 | 1532 | 4800 | 200 | 38 |
| **Average** | **460** | | **607** | | **33** |

TABLE VI
IMPROVEMENT IN NUMBER OF TASKS SCHEDULED FOR NON-UNIFORM LAXITY APPROACH COMPARED TO CONVENTIONAL EDF -- SIMULATED ON SESC

| Number of tasks | EDF | | Non-uniform laxity applied to EDF | | Improvement (%) |
|---|---|---|---|---|---|
| | Tasks scheduled | Tasks missed | Tasks scheduled | Tasks missed | |
| 8 | 3 | 5 | 4 | 4 | 33 |
| 15 | 8 | 7 | 13 | 2 | 63 |
| 20 | 11 | 9 | 15 | 5 | 36 |
| 30 | 20 | 10 | 27 | 3 | 35 |
| 45 | 30 | 15 | 40 | 5 | 33 |
| 60 | 45 | 15 | 59 | 1 | 31 |
| 75 | 50 | 25 | 69 | 6 | 38 |
| 80 | 55 | 25 | 75 | 5 | 36 |
| 90 | 65 | 25 | 87 | 3 | 34 |
| 100 | 70 | 30 | 95 | 5 | 36 |
| 200 | 125 | 75 | 174 | 26 | 39 |
| 500 | 345 | 155 | 475 | 25 | 38 |
| 700 | 490 | 210 | 650 | 50 | 33 |
| 900 | 636 | 264 | 850 | 50 | 34 |
| 1000 | 690 | 310 | 920 | 80 | 33 |
| 2000 | 1491 | 509 | 1967 | 33 | 32 |
| 5000 | 3658 | 1342 | 4985 | 15 | 36 |
| **Average** | **458** | | **618** | | **36** |

The results clearly show that EDF modified by non-uniform laxity improves both task utilisation and task schedulability significantly.

# IV CONCLUSION AND FUTURE WORK

In this paper, a new scheduler applying non-uniform laxity to EDF for aperiodic tasks has been proposed. The algorithm has been tested for upto 5000 random task sets and upto 100 cores. The scheduler improves task utilisation by 35% and increases the number of tasks being scheduled by 36%, compared to conventional EDF.

Other slack time measures like critical scaling and skewness can also be explored in future to improve task utilisation.